\documentclass[11pt]{article}

\usepackage[final]{acl}

\usepackage{times}
\usepackage{latexsym}

\usepackage[T1]{fontenc}

\usepackage[utf8]{inputenc}

\usepackage{microtype}

\usepackage{inconsolata}

\usepackage{amsmath,amssymb,amsfonts}
\usepackage{graphicx}
\usepackage{booktabs}
\usepackage{tabularx}
\usepackage{multirow}
\usepackage{enumitem}
\usepackage{array}
\usepackage{float}
\usepackage{pifont}
\usepackage{placeins}
\usepackage[table]{xcolor}

\newcommand{\cmark}{\ding{51}} 
\newcommand{\xmark}{\ding{55}} 

\definecolor{highlight}{HTML}{A9D3CB}

\title{A Study of Reinforcement Learning Reward for Code Generation}

\author{ 
Shuyin Ouyang\\ 
King's College London, UK \\
\texttt{shuyin.ouyang@kcl.ac.uk} \And 
Zhaozhi Qian\\ 
Elm Company, UK \\
\texttt{zqian@elm.sa} \And 
Faroq AL-Tam\\
Elm Company, KSA  \\
\texttt{faltam@elm.sa} \AND 
Muhammad AL-Qurishi\\
Elm Company, KSA \\
\texttt{mualqurishi@elm.sa} \And 
Jie M. Zhang\\
King's College London, UK \\
\texttt{jie.zhang@kcl.ac.uk}
}

\begin{document}
\maketitle
\begin{abstract}
Reinforcement learning (RL) is widely used to improve code generation, typically with rewards based on execution-based test outcomes, but such rewards can be slow, sparse, and misleading, destabilizing training.
Execution-free rewards, such as parser- or similarity-based rewards, have been proposed, yet it remains unclear how their effectiveness varies across different model architectures, particularly as diffusion language models emerge as an alternative to autoregressive (AR) models.
We present a comprehensive study comparing five reward functions, applied individually, in combination, and via progressive scheduling, across two AR models and two diffusion models, evaluated on HumanEval(+), MBPP(+), and LiveCodeBench.
In the single-reward setting, the static checking reward is a strong alternative to the execution-based reward, improving performance in three of four models and reducing wall-clock training time by up to 50\% for some AR models.
These findings suggest that developers should also consider lightweight yet effective alternatives to expensive execution-based rewards in RL.
Our code is available.\footnote{\url{https://github.com/ShuyinOuyang/RL-Reward-Code}}

\end{abstract}

\section{Introduction}
\label{section: Introduction}

Large language models (LLMs) have substantially advanced code generation, with both autoregressive (AR) models and diffusion language models showing strong potential for generating code from natural-language specifications~\cite{chen2021evaluating, gong2025diffucoder}.
AR models generate code sequentially from left to right, while diffusion language models generate or refine code through iterative denoising, offering a different generation paradigm~\cite{gong2025diffucoder}.
Despite their architectural differences, both types of models face the same core challenge: high-quality code generation requires more than producing plausible text~\cite{liu2026beyond}.
A generated solution must follow the expected output format, be syntactically valid, satisfy static constraints, use appropriate code structures, and ultimately pass execution-based tests.
These requirements make code generation a particularly suitable domain for reinforcement learning (RL)~\cite{zhao2026d1}, because many aspects of code quality can be automatically checked.

Execution-based rewards~\cite{le2022coderl} directly reflect functional correctness in code generation and avoid the need for expensive human annotation, making them popular for large-scale RL training.
However, such rewards are often sparse and delayed: many generated programs receive zero reward, especially in the early stage of training, even when they are partially correct or close to a valid solution~\cite{le2022coderl, dou2024stepcoder}.
As a result, RL training may suffer from weak learning signals, unstable optimization, and inefficient exploration.
This limitation raises an important question: whether additional or alternative reward designs can provide more informative supervision and improve RL training for code generation.

This uncertainty becomes more important as code generation models diversify.
Because AR models and diffusion models use fundamentally different approaches to modeling language and code, the same reward may have different effects on training stability, exploration, and final correctness.
However, there is still limited empirical understanding of whether reward signals designed for AR code models transfer effectively to diffusion code models.

In this paper, we conduct a series of experiments using both AR-based and diffusion-based language models, spanning five reward functions (format, syntax, static checking, similarity, and execution) applied individually, in combination, and under a progressive scheduling regime~\cite{ye2025process}.
We perform RL training on AceCode-87k~\cite{zeng2025acecoder}, and evaluate the resulting models on HumanEval, HumanEval+, MBPP, MBPP+, and LiveCodeBench~\cite{chen2021evaluating, liu2023your, austin2021program, jain2024livecodebench}.

Our results show that while execution reward delivers the  largest performance gain for Qwen2.5-Coder, static checking proves to be the consistently superior alternative across the remaining three models. In contrast, format and similarity rewards frequently fail to beat the unrewarded baseline.
Furthermore, while execution-based rewards impose a steep wall-clock penalty, nearly doubling training time for AR models like Qwen2.5-Coder and Seed-Coder, static checking eliminates this bottleneck to deliver up to a 50\% speedup.
In multi-reward settings, diffusion models are highly sensitive to reward interaction: naive reward combination significantly degrades performance for models like DiffuCoder, even when individual components provide positive signals. 
Progressive scheduling successfully mitigates this interference, avoiding the failure mode and enabling the model to surpass the unrewarded baseline across most benchmarks.

To summarize, this paper makes the following contributions:
\begin{itemize}
    \item We present the first study of reward function design for RL-based code generation across both AR and diffusion language models, covering five reward types.
    \item We identify static-checking reward as a strong alternative to execution-based reward in the single-reward setting, providing performance improvements for both model types.
    \item We show that reward combination and progressive scheduling affect training stability differently by architecture, and that progressive reward introduction substantially mitigates the training collapse observed for diffusion models under naive combination.
    \item We characterize the training speed and learning dynamics of different reward signals across architectures, and show that execution-free rewards lead to faster training time and earlier convergence.
\end{itemize}

\section{Related Work}
\label{section: Related Work}

\paragraph{Code Language Models with Reinforcement Learning.}
Code LLMs achieve strong performance through large-scale pre-training and instruction tuning~\cite{hui2024qwen2, seed2025seed}, yet they still struggle on real-world code generation and repair~\cite{ouyang2026dscodebench, ouyang2025knowledge}. 
RL is increasingly used as a post-training step, typically sampling complete programs and updating the policy with compiler or unit-test feedback~\cite{dou2024stepcoder, shojaee2023execution}.
Diffusion language models offer an alternative to left-to-right decoding, iteratively denoising a masked sequence and revising multiple positions at once. 
DiffuCoder adapts masked diffusion modeling to code with a diffusion-native RL method, coupled-GRPO~\cite{gong2025diffucoder}, and Dream-Coder applies RL with verifiable rewards to a discrete diffusion code model~\cite{xie2025dream}.
Both works show RL benefits AR and diffusion code models alike, but focus on model design rather than reward design; we instead hold the reward function as the primary variable and compare its effect across both architectures.

\paragraph{Reward Functions for Code Generation.}
Execution correctness is the dominant reward in recent practice~\cite{le2022coderl, liu2023rltf, dou2024stepcoder}, but it is coarse: unparseable code and code that fails all tests receive the same zero reward.
Prior work seeks denser or execution-free alternatives: compiler feedback distinguishes syntax, type, and runtime errors~\cite{shojaee2023execution, dou2024stepcoder}; unit-test feedback can be decomposed by pass count~\cite{liu2023rltf}; and static analysis rewards score code quality without execution~\cite{yao2026training}, though combining signals can induce reward hacking~\cite{fu2025reward}.
We organize these along a spectrum, \textit{format}, \textit{syntax}, \textit{static-checking}, \textit{similarity}, and \textit{execution} unlike prior single-algorithm or model-specific studies, and compare them under one unified setting across both AR and diffusion code models.

\section{Experiment Design}
\label{section: Experiment Design}

This study is organized around the following experiments.

\textbf{Experiment 1: How effective is each single reward for RL-based code generation?}
We examine how five reward functions, \textit{format}, \textit{syntax}, \textit{static checking}, \textit{similarity}, and \textit{execution}, each affect RL training when used in isolation.

\textbf{Experiment 2: How do different reward combinations affect RL training?}
Since RL training in practice often combines multiple reward signals rather than relying on a single one, we study how different combinations of the five rewards affect training relative to their individual counterparts in Experiment 1.

\textbf{Experiment 3: Does progressive reward scheduling improve RL training?}
Rather than applying combined rewards simultaneously throughout training, we consider introducing reward components progressively over the training period, and evaluate how this scheduling changes training outcomes relative to the simultaneous combinations in Experiment 2.

\textbf{Experiment 4: What are the training time and learning dynamics of different reward designs for AR and diffusion code models?}
Because RL training is computationally expensive, we further analyze the training-time and learning dynamics associated with each reward, and how these costs differ between AR and diffusion code models.

\subsection{Model Selection}
\label{subsection: Model Selection}

We evaluate four base code language models spanning two generation paradigms: two AR models, \textbf{Qwen2.5-Coder-7B-Instruct}~\cite{hui2024qwen2} and \textbf{Seed-Coder-8B-Instruct}~\cite{seed2025seed}, and two diffusion language models (DLMs), \textbf{DiffuCoder-7B-Instruct}~\cite{gong2025diffucoder} and \textbf{Dream-Coder-v0-Instruct-7B}~\cite{xie2025dream}. All four models are specialized for code generation and are comparable in scale (7--8B parameters), allowing us to contrast generation paradigm (AR vs.\ diffusion) while limiting confounds from model size.

\subsection{Training}
\label{subsection: Training}

We perform RL training on AceCode-87k~\cite{zeng2025acecoder}.
In this paper, we focus on the hard partition because execution-based reward is most sparse in this regime, making it the setting where denser reward alternatives are most likely to provide additional benefit and where reward-interference effects are most likely to surface.

Models are trained with Group Relative Policy Optimization (GRPO)~\cite{shao2024deepseekmath}, using the five reward functions, format, syntax, static checking, similarity, and execution, applied under three regimes: individually (Experiment 1), in simultaneous combination (Experiment 2), and under a progressive scheduling regime (Experiment 3).
For combined-reward runs, the component rewards are summed into a single scalar signal used for the GRPO advantage estimate.
For progressive-scheduling runs, reward components are introduced in stages (format  → syntax → static checking → execution), with the model advancing to the next stage only once its rolling performance on the current stage's reward crosses a fixed threshold (0.5 or 0.8).

Each reward function is normalized to the same scale before being incorporated into training, so that no single component dominates the combined signal solely due to differences in raw magnitude.
Following the cost tiers used in Experiment 4, format and syntax rewards are classified as \textit{lightweight}, since both are computed directly from generated text or its parsed AST without any external analysis;
static checking and similarity are classified as \textit{static analysis}, since both require an additional analysis or comparison step but do not execute the generated code; 
and execution is classified as \textit{execution-based}, since it requires running the generated code in a sandboxed executor and is substantially more expensive per rollout than the other four rewards. 
The cost is examined directly in Experiment 4.

Each run uses matched hyperparameters and hardware across models and reward configurations to ensure a fair comparison.
Training proceeds for up to 719 steps per run, with runtime, throughput, and reward statistics logged at each step.
\textit{Full details on training are given in Appendix~\ref{appendix: Training}.}

\begin{table*}[t]
\caption{Experiment 1: Effect of individual reward components on code generation benchmarks. The base models (grey rows) are shown as ``--'' for reference. \colorbox{highlight}{Light green} cells indicate a reward outperforming the base model on that metric; \textbf{bold} indicates the best value (not including the baseline) in each column, within each model block.}
\centering
\setlength{\aboverulesep}{0pt}
\setlength{\belowrulesep}{0pt}

\resizebox{0.9\linewidth}{!}{%
\label{tab:RQ1_only_reward_results}
\begin{tabular}{@{}llrrrrr}
\toprule
\textbf{Model} & \textbf{Reward} & \textbf{HumanEval} & \textbf{HumanEval+} & \textbf{MBPP} & \textbf{MBPP+} & \textbf{LiveCodeBench} \\
\midrule
\multirow{6}{*}{Qwen2.5-Coder}
 & \cellcolor{gray!20}--               & \cellcolor{gray!20}0.884 & \cellcolor{gray!20}0.841 & \cellcolor{gray!20}0.825 & \cellcolor{gray!20}0.709 & \cellcolor{gray!20}0.297 \\
\cmidrule{2-7}
 & Format           & 0.884 & 0.811 & 0.825 & 0.704 & \cellcolor{highlight}0.298 \\
 & Syntax           & \cellcolor{highlight}0.896 & 0.817 & \cellcolor{highlight}0.844 & \cellcolor{highlight}0.717 & 0.295 \\
 & Similarity       & 0.866 & 0.799 & \cellcolor{highlight}0.839 & 0.709 & 0.280 \\
 & Static Checking  & 0.854 & 0.787 & \cellcolor{highlight}\textbf{0.862} & 0.698 & \cellcolor{highlight}\textbf{0.300} \\
 & Execution        & \cellcolor{highlight}\textbf{0.915} & \cellcolor{highlight}\textbf{0.854} & \cellcolor{highlight}0.852 & \cellcolor{highlight}\textbf{0.720} & 0.291 \\
\cmidrule{1-7}
\multirow{6}{*}{Seed-Coder}
 & \cellcolor{gray!20}--               & \cellcolor{gray!20}0.848 & \cellcolor{gray!20}0.787 & \cellcolor{gray!20}0.852 & \cellcolor{gray!20}0.712 & \cellcolor{gray!20}0.334 \\
\cmidrule{2-7}
 & Format           & 0.841 & 0.774 & \cellcolor{highlight}0.860 & 0.709 & \cellcolor{highlight}\textbf{0.341} \\
 & Syntax           & 0.841 & 0.780 & \cellcolor{highlight}0.860 & 0.712 & 0.332 \\
 & Similarity       & 0.841 & 0.780 & \cellcolor{highlight}\textbf{0.868} & \cellcolor{highlight}\textbf{0.730} & \cellcolor{highlight}0.336 \\
 & Static Checking  & \cellcolor{highlight}\textbf{0.854} & \cellcolor{highlight}0.793 & \cellcolor{highlight}\textbf{0.868} & 0.709 & \cellcolor{highlight}0.337 \\
 & Execution        & 0.848 & \cellcolor{highlight}\textbf{0.799} & \cellcolor{highlight}0.862 & 0.709 & \cellcolor{highlight}0.336 \\
\cmidrule{1-7}
\multirow{6}{*}{DiffuCoder}
 & \cellcolor{gray!20}--               & \cellcolor{gray!20}0.652 & \cellcolor{gray!20}0.616 & \cellcolor{gray!20}0.701 & \cellcolor{gray!20}0.590 & \cellcolor{gray!20}0.134 \\
\cmidrule{2-7}
 & Format           & 0.622 & 0.567 & \cellcolor{highlight}0.706 & \cellcolor{highlight}0.595 & 0.126 \\
 & Syntax           & 0.494 & 0.445 & 0.651 & 0.561 & 0.100 \\
 & Similarity       & \cellcolor{highlight}\textbf{0.689} & \cellcolor{highlight}\textbf{0.634} & 0.688 & 0.577 & 0.071 \\
 & Static Checking  & \cellcolor{highlight}0.677 & 0.610 & \cellcolor{highlight}\textbf{0.725} & \cellcolor{highlight}0.608 & \cellcolor{highlight}\textbf{0.155} \\
 & Execution        & 0.591 & 0.549 & \cellcolor{highlight}0.720 & \cellcolor{highlight}\textbf{0.627} & 0.123 \\
\cmidrule{1-7}
\multirow{6}{*}{Dream-Coder}
 & \cellcolor{gray!20}--               & \cellcolor{gray!20}0.701 & \cellcolor{gray!20}0.646 & \cellcolor{gray!20}0.765 & \cellcolor{gray!20}0.656 & \cellcolor{gray!20}0.174 \\
\cmidrule{2-7}
 & Format           & 0.640 & 0.598 & 0.765 & 0.646 & \cellcolor{highlight}0.182 \\
 & Syntax           & 0.665 & 0.622 & \cellcolor{highlight}\textbf{0.794} & \cellcolor{highlight}\textbf{0.683} & 0.167 \\
 & Similarity       & 0.640 & 0.598 & \cellcolor{highlight}0.770 & \cellcolor{highlight}0.669 & \cellcolor{highlight}0.188 \\
 & Static Checking  & \cellcolor{highlight}\textbf{0.738} & \cellcolor{highlight}\textbf{0.689} & \cellcolor{highlight}0.786 & \cellcolor{highlight}0.667 & \cellcolor{highlight}0.189 \\
 & Execution        & 0.683 & 0.640 & \cellcolor{highlight}0.780 & \cellcolor{highlight}0.669 & \cellcolor{highlight}\textbf{0.199} \\
\bottomrule
\end{tabular}
}

\end{table*}
\vspace*{-0.1cm}

\subsection{Evaluation}
\label{subsection: Evaluation}

We evaluate all trained models on three widely used code generation benchmarks: \textbf{HumanEval(+)}, \textbf{MBPP(+)}, and \textbf{LiveCodeBench}.
HumanEval and MBPP (together with their ``+''-augmented variants, which extend the original test suites with additional edge-case tests) measure functional correctness on relatively short, self-contained programming problems, while LiveCodeBench (v6) evaluates performance on more recent and diverse problems designed to reduce train--test contamination.
We report pass@1 accuracy for all benchmarks under greedy decoding, which avoids the output non-determinism of sampling-based LLM code generation~\cite{ouyang2025empirical}.
In addition, to assess training cost, we log mean training time (in hours) and throughput (samples/s) for each model and reward configuration over the course of training, allowing us to relate benchmark gains to the compute required to obtain them.
\textit{Full details on evaluation are given in Appendix~\ref{appendix: Evaluation}.}

\section{Results}
\label{section: Results}

\begin{table*}[h!]
\centering
\small
\setlength{\tabcolsep}{6pt}
\setlength{\extrarowheight}{2pt}
\setlength{\aboverulesep}{0pt}
\setlength{\belowrulesep}{0pt}
\caption{Experiment 2: Effect of combined reward components on code generation benchmarks (hard-difficulty runs). Reward abbreviations: F = format, Sy = syntax, St = static checking, Si = similarity, E = execution.}
\label{tab:RQ2_reward_combinations}
\begin{tabular}{@{}llrrrrr}
\toprule
\textbf{Model} & \textbf{Reward Combination} & \textbf{HumanEval} & \textbf{HumanEval+} & \textbf{MBPP} & \textbf{MBPP+} & \textbf{LiveCodeBench} \\
\midrule
\multirow{5}{*}{Qwen2.5-Coder}
 & \cellcolor{gray!20}--                  & \cellcolor{gray!20}0.884 & \cellcolor{gray!20}0.841 & \cellcolor{gray!20}0.825 & \cellcolor{gray!20}0.709 & \cellcolor{gray!20}0.297 \\
\cmidrule{2-7}
 & F + E                & 0.866 & \textbf{0.835} & \cellcolor{highlight}0.849 & \cellcolor{highlight}0.720 & 0.291 \\
 & F + E + Sy           & 0.884 & 0.823 & \cellcolor{highlight}\textbf{0.857} & \cellcolor{highlight}\textbf{0.738} & \cellcolor{highlight}0.307 \\
 & F + E  + Sy + St     & 0.872 & 0.805 & \cellcolor{highlight}\textbf{0.857} & \cellcolor{highlight}0.712 & \cellcolor{highlight}\textbf{0.308} \\
 & F + E  + Sy + St + Si & \cellcolor{highlight}\textbf{0.909} & \textbf{0.835} & \cellcolor{highlight}0.854 & 0.709 & 0.297 \\
\cmidrule{1-7}
\multirow{5}{*}{DiffuCoder}
 & \cellcolor{gray!20}--                  & \cellcolor{gray!20}0.652 & \cellcolor{gray!20}0.616 & \cellcolor{gray!20}0.701 & \cellcolor{gray!20}0.590 & \cellcolor{gray!20}0.134 \\
\cmidrule{2-7}
 & F + E                & 0.476 & 0.427 & \cellcolor{highlight}0.704 & \cellcolor{highlight}\textbf{0.616} & 0.095 \\
 & F + E + Sy          & 0.488 & 0.457 & 0.696 & \cellcolor{highlight}0.598 & 0.086 \\
 & F + E + Sy + St      & 0.573 & 0.530 & \cellcolor{highlight}\textbf{0.725} & \cellcolor{highlight}0.608 & 0.065 \\
 & F + E + Sy + St + Si & \textbf{0.585} & \textbf{0.537} & \cellcolor{highlight}0.709 & \cellcolor{highlight}{0.606} & \textbf{0.096} \\
\bottomrule
\end{tabular}
\end{table*}

\subsection{Experiment 1: Single Reward}

Table~\ref{tab:RQ1_only_reward_results} reports pass@1 accuracy under each reward across all four models in all the benchmarks.
For Qwen2.5-Coder, execution reward gives the largest gains, raising HumanEval from 0.884 to 0.915, and tops three of five benchmarks.
For Seed-Coder, execution reward is weaker, offering only a marginal improvement (e.g., HumanEval+), while static checking performs better overall, reaching its highest scores on HumanEval and MBPP.
This contrast suggests execution-based supervision is not always the strongest option, even between two AR models of similar scale.

Across all four models, execution and static checking account for most of the best-per-benchmark results.
Static checking gives DiffuCoder its best scores on MBPP and LiveCodeBench, and gives Dream-Coder its best HumanEval and HumanEval+ scores as well. 
This suggests rewards tied to code correctness, whether checked through execution or static analysis, are more often associated with gains than rewards based on surface features.

Format and similarity rewards perform more weakly and inconsistently.
Format reward slightly lowers Qwen2.5-Coder's HumanEval+ and MBPP+ relative to baseline, and similarity reward lowers its HumanEval score as well.
For DiffuCoder, similarity reward lowers LiveCodeBench, from 0.134 to 0.071, despite modestly raising HumanEval.
These patterns suggest that rewards based on surface-level or lexical similarity can improve one benchmark while degrading another, rather than yielding consistent gains.

Syntax reward produces the clearest architecture-level difference.
For the two AR models, syntax reward changes scores slightly, with small gains on Qwen2.5-Coder's HumanEval and Seed-Coder's MBPP.
For DiffuCoder, the same reward lowers every benchmark score, most sharply on HumanEval (0.652 to 0.494) and LiveCodeBench (0.134 to 0.100), suggesting its non-AR generation process may be more susceptible to reward hacking~\cite{wang2026reward} under a narrow, easily satisfied signal.

\textbf{Summary:} While no reward strategy is universally optimal, static checking emerges as a compelling alternative to execution-based reward. 
Although execution reward achieves the highest peak performance for Qwen2.5-Coder, static checking outperforms it across the other three models—offering practitioners a far more reliable default, especially given that format and similarity rewards frequently trade gains on one benchmark for losses on another.
In addition, syntax reward maintains stability for autoregressive models but severely degrades performance across every benchmark for DiffuCoder. 
Practitioners should exercise caution when applying syntax-based rewards to diffusion architectures, as they carry a high risk of catastrophic performance loss.

\subsection{Experiment 2: Combined Reward}

Table~\ref{tab:RQ2_reward_combinations} shows that reward combination affects the two models in largely opposite ways.
For Qwen2.5-Coder, the combination is broadly beneficial: the full five-reward combination reaches 0.909 on HumanEval, above the base model's 0.884, though it still trails the single best reward, execution alone at 0.915.
This suggests that combining signals can match or exceed several individual rewards, even though it does not surpass the single strongest one.

No combination dominates across all benchmarks, however, pointing to a trade-off rather than a uniform gain.
The three- and four-reward combinations reach the best MBPP+ and LiveCodeBench scores among combination runs, but at the cost of the lowest HumanEval+ scores, falling to 0.805.
HumanEval+ only reaches 0.835 once similarity is added in the full combination, suggesting partial combinations can over-optimize some benchmarks at the expense of others.

For DiffuCoder, combination behaves quite differently: every combination underperforms the base model on HumanEval, HumanEval+, and LiveCodeBench, with the two-reward combination dropping HumanEval from 0.652 to 0.476.
This contrasts with Experiment 1, where individual rewards such as static checking matched or exceeded the base model, and no combination run in this experiment approaches those single-reward results.
This gap suggests that for DiffuCoder, combining rewards may introduce interference rather than complementary learning.

Within DiffuCoder's results, performance does not decline steadily as more rewards are added: it drops sharply under the two-reward combination (0.476), then partly recovers to 0.585 on HumanEval, as static checking and similarity are layered in.
This pattern suggests static checking and similarity may act as partial correctives to more disruptive rewards like format and syntax, though the full combination still falls short of baseline for this model.

\textbf{Summary:} These results indicate that reward combination tends to help the AR model tested here while consistently hurting the diffusion model, reinforcing the architecture-dependent sensitivity observed in Experiment 1.
The non-monotonic recovery seen as more rewards are layered into DiffuCoder's combination also hints that which rewards are added together, not just how many, may influence training stability, a question taken up further in Experiment 3.

\begin{table*}[t]
\centering
\small
\setlength{\tabcolsep}{6pt}
\setlength{\extrarowheight}{2pt}
\setlength{\aboverulesep}{0pt}
\setlength{\belowrulesep}{0pt}
\caption{Experiment 3: Effect of progressive reward scheduling on code generation benchmarks. Reward abbreviations: F = format, Sy = syntax, St = static checking, E = execution.}
\label{tab:RQ3_threshold_results}
\begin{tabular}{@{}llcrrrrr}
\toprule
\textbf{Model} & \textbf{Reward Schedule} & \textbf{Thresh.} & \textbf{HumanEval} & \textbf{HumanEval+} & \textbf{MBPP} & \textbf{MBPP+} & \textbf{LiveCodeBench} \\
\midrule
\multirow{5}{*}{Qwen2.5-Coder}
 & \cellcolor{gray!20}--                     & \cellcolor{gray!20}--  & \cellcolor{gray!20}0.884 & \cellcolor{gray!20}0.841 & \cellcolor{gray!20}0.825 & \cellcolor{gray!20}0.709 & \cellcolor{gray!20}0.297 \\
\cmidrule{2-8}
 & F $\rightarrow$ Sy $\rightarrow$ E            & 0.5 & \cellcolor{highlight}0.896 & \cellcolor{highlight}\textbf{0.848} & \cellcolor{highlight}0.836 & \cellcolor{highlight}0.712 & \cellcolor{highlight}0.302 \\
 & F $\rightarrow$ Sy $\rightarrow$ E            & 0.8 & \cellcolor{highlight}\textbf{0.902} & 0.829 & \cellcolor{highlight}0.847 & \cellcolor{highlight}0.717 & \cellcolor{highlight}0.303 \\
 & F $\rightarrow$ Sy $\rightarrow$ St $\rightarrow$ E & 0.5 & 0.884 & 0.823 & \cellcolor{highlight}\textbf{0.857} & \cellcolor{highlight}0.728 & \cellcolor{highlight}\textbf{0.304} \\
 & F $\rightarrow$ Sy $\rightarrow$ St $\rightarrow$ E & 0.8 & \cellcolor{highlight}\textbf{0.902} & 0.835 & \cellcolor{highlight}\textbf{0.857} & \cellcolor{highlight}\textbf{0.733} & \cellcolor{highlight}0.303\\
\cmidrule{1-8}
\multirow{5}{*}{DiffuCoder}
 & \cellcolor{gray!20}--                     & \cellcolor{gray!20}--  & \cellcolor{gray!20}0.652 & \cellcolor{gray!20}0.616 & \cellcolor{gray!20}0.701 & \cellcolor{gray!20}0.590 & \cellcolor{gray!20}0.134 \\
\cmidrule{2-8}
 & F $\rightarrow$ Sy $\rightarrow$ E            & 0.5 & 0.585 & 0.549 & \cellcolor{highlight}0.709 & \cellcolor{highlight}0.606 & 0.105 \\
 & F $\rightarrow$ Sy $\rightarrow$ E            & 0.8 & 0.616 & 0.573 & \cellcolor{highlight}0.714 & \cellcolor{highlight}\textbf{0.624} & 0.118 \\
 & F $\rightarrow$ Sy $\rightarrow$ St $\rightarrow$ E & 0.5 & 0.634 & \textbf{0.585} & \cellcolor{highlight}0.717 & \cellcolor{highlight}0.616 & 0.114 \\
 & F $\rightarrow$ Sy $\rightarrow$ St $\rightarrow$ E & 0.8 & \textbf{0.646} & 0.573 & \cellcolor{highlight}\textbf{0.725} & \cellcolor{highlight}0.616 & \cellcolor{highlight}\textbf{0.138} \\
\bottomrule
\end{tabular}
\end{table*}

\subsection{Experiment 3: Progressive Reward}

Table~\ref{tab:RQ3_threshold_results} replaces the simultaneous reward combination used in Experiment 2 with a progressive schedule, where reward components are introduced in stages, and the model advances only once its performance on the current stage crosses a threshold.
Comparing matched reward sets across Table~\ref{tab:RQ2_reward_combinations} and Table~\ref{tab:RQ3_threshold_results} shows that progressive reward helps in nearly every case tested.
For Qwen2.5-Coder, the progressive schedule improves across most metrics for both the three- and four-component reward sets, with LiveCodeBench the only benchmark that stays roughly flat.

The effect is more pronounced for DiffuCoder, where Experiment 2 showed simultaneous combination actively hurt performance.
Under progressive scheduling, both the three- and four-component sets improve on nearly every benchmark relative to their combined-reward counterparts, with the largest gains on HumanEval and LiveCodeBench.
This suggests progressive scheduling can meaningfully ease the reward-interference problem identified in Experiment 2, rather than simply stacking rewards at once.

The best progressive configuration for DiffuCoder, the four-component set at threshold 0.8, exceeds the unrewarded base model on MBPP, MBPP+, and LiveCodeBench, a contrast to every combination run in Experiment 2, most of which fell below baseline on LiveCodeBench and other benchmarks.
HumanEval and HumanEval+ remain slightly below baseline under this setting, but the gap narrows substantially relative to simultaneous combination.
This suggests that for reward-sensitive architectures like DiffuCoder, the order and pace of reward introduction may matter as much as reward choice itself.

Raising the advancement threshold from 0.5 to 0.8 generally improves final performance, most consistently on HumanEval and LiveCodeBench, with the effect somewhat larger for DiffuCoder than for Qwen2.5-Coder.
HumanEval+ is a partial exception, with mixed results across settings. 
This suggests that requiring stronger performance at each stage before moving on tends to produce more stable downstream training, though not for every benchmark.

\textbf{Summary:} These results suggest progressive scheduling is not just an incremental refinement of combination but a meaningfully different strategy for reward-sensitive architectures, largely recovering what simultaneous combination lost for DiffuCoder while offering smaller, steadier gains for Qwen2.5-Coder.

\subsection{Experiment 4: Training Efficiency}

\begin{figure}[h!]
\centering
\includegraphics[width=\linewidth, trim=0 5 0 5, clip]{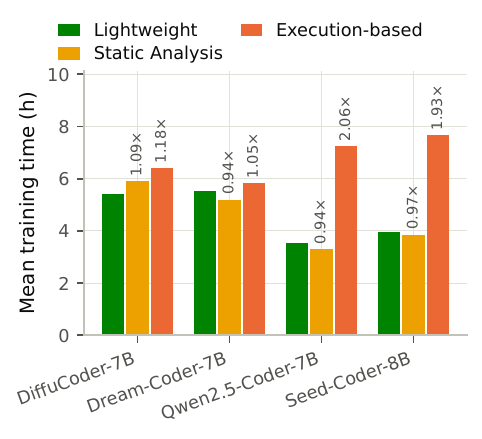}
\caption{Experiment 4: Mean training time per reward-function cost tier, by model. 
Rewards are grouped into three tiers: \textit{lightweight} (format, syntax), \textit{static analysis} (static checking, similarity), and \textit{execution-based} (execution). Annotated multipliers give each tier's overhead relative to the lightweight baseline for that model.}
\label{fig:rq4-overhead}
\end{figure}

Figure~\ref{fig:rq4-overhead} shows that execution-based reward adds substantial runtime overhead relative to the lightweight baseline for the two AR models, roughly doubling training time for Qwen2.5-Coder (2.06×) and Seed-Coder (1.93×), while the static analysis tier stays close to baseline cost for both.
This suggests static analysis or reference-similarity comparison costs about as much wall-clock time as simple text or AST checks, since neither requires executing the candidate code.

This overhead is far from uniform across architectures.
For the two diffusion models, execution-based reward adds only a mild premium relative to their own lightweight baseline (0.94× to 1.09× across tiers), and throughput falls only modestly compared to the AR models. 
However, diffusion rollouts are already substantially more costly than AR rollouts in absolute terms, so this smaller relative overhead still runs on top of a higher baseline cost per rollout.
This gap suggests the wall-clock cost of richer reward tiers depends heavily on how each architecture generates text, and that reward-tier overhead should be read as a relative, not absolute, comparison across architectures.

\begin{figure}[h!]
\centering
\includegraphics[width=\linewidth, trim=0 5 0 5, clip]{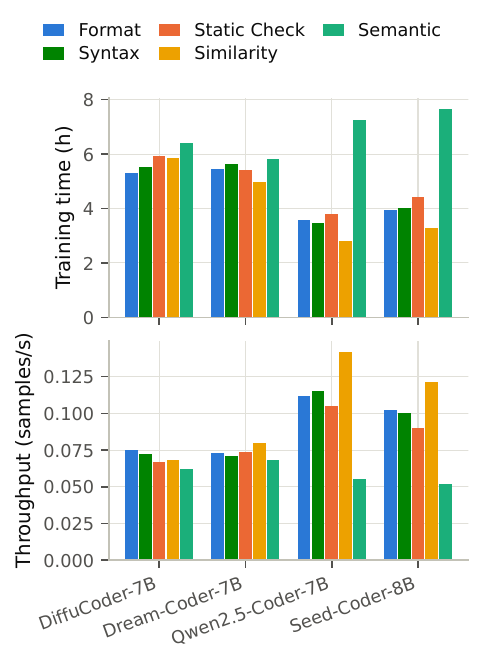}
\caption{Experiment 4: Training time (top) and throughput in samples/s (bottom) by model and reward function.}
\label{fig:rq4-time-speed}
\end{figure}

Figure~\ref{fig:rq4-time-speed} shows that similarity reward is one of the fastest configurations for the AR models, since comparing against a concise reference solution tends to produce shorter completions than either lightweight reward.
The two diffusion models show a different pattern: their training time and throughput stay comparatively stable across reward functions, consistent with diffusion decoding running a fixed denoising budget that the reward signal does not readily lengthen or shorten.
This suggests that for AR models, completion length itself is reward-shaped and compounds with sandbox latency, while for diffusion models the added cost of richer rewards is mainly the sandbox round-trip itself.

\begin{figure}[h!]
\centering
\includegraphics[width=\linewidth, trim=0 5 0 5, clip]{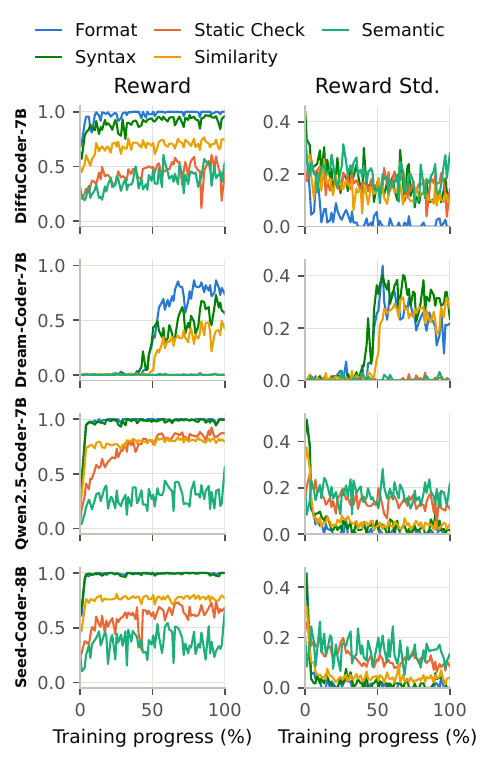}
\caption{Experiment 4: Reward (left) and reward standard deviation (right) over training
progress, by model and reward function.}
\label{fig:rq4-reward-curves}
\end{figure}

Figure~\ref{fig:rq4-reward-curves} shows that 
execution-free rewards usually converge faster than the execution-based semantic reward; this can be seen both in the earlier plateauing of execution-free rewards and the smaller reward standard deviation for them. Due to the faster convergence, they are well-suited to the initial phase of the progressive curriculum. 
Among the execution-free rewards, the format reward is the one that converges the fastest, though this likely reflects superficial compliance rather than genuine correctness, since reward under the static analysis and execution-based tiers stays well below 1.0 for every model. 

Across the two model types, reward standard deviation late in training is generally higher for the diffusion models than the AR models, indicating a persistently noisier learning signal well past the point where AR runs have largely settled. 

\textbf{Summary:} In characterizing the training efficiency and learning dynamics of different reward signals, we find that execution-free rewards, most notably static checking, offer two distinct advantages over execution-based rewards.
First, they eliminate sandboxing overhead, cutting wall-clock training time by up to 50\% for autoregressive models.
Second, they exhibit faster convergence dynamics, characterized by earlier reward plateaus and lower variance, which makes execution-free signals uniquely well-suited for the initial phase of a progressive curriculum.

\vspace*{-0.2cm}


\section{Discussion}
\label{section: Discussion}

The results in Section~\ref{section: Results} point to reward effectiveness being closely tied to model architecture rather than to any single reward function in isolation.
We step back from the individual experiments to consider what might explain this architecture dependence, how our findings relate to prior work on reward hacking, and what they imply for practitioners and for RL algorithm design more broadly.

\paragraph{Why are diffusion models more reward-sensitive?}
A plausible explanation lies in generation flexibility.
Experiment 4 shows that AR completion length grows under richer reward tiers, while diffusion completion length stays roughly fixed, since diffusion decoding runs a constant denoising budget regardless of the reward signal~\cite{rossi2026diffusion, li2025beyond}.
AR models can therefore absorb reward pressure gradually through output length, while diffusion models must express any behavioral change within a fixed-length process, which may concentrate reward pressure on discrete token-level choices, a more brittle optimization target.
This account is consistent with prior evidence that combining reward signals can induce reward hacking even when individual signals are well-behaved~\cite{fu2025reward}.
Most such evidence comes from AR settings, but our results suggest the same risk arises for diffusion models at a larger magnitude and under weaker conditions: a single narrow reward (syntax) is already sufficient to destabilize DiffuCoder, whereas our AR models remain stable under the same reward in isolation. 
We treat this length-flexibility account as a hypothesis motivated by our results rather than one we directly test, and see generation-level diagnostics as a natural way to examine it in future work.

\paragraph{Practical implications.}
For AR code models, execution and static-checking rewards are reasonable default building blocks, while format and similarity rewards are better used as auxiliary signals than primary objectives, since they frequently underperform the unrewarded baseline alone.
Combination is worth attempting for AR models, but more rewards are not always better: partial combinations can over-optimize specific benchmarks at the expense of others, and in our study only the full five-reward combination rebalanced these pressures.

For diffusion code models, naively summing multiple reward terms is a high-risk default that can erase the gains available from any one reward used alone.
Practitioners should prefer staged reward introduction, moving from denser, easier-to-satisfy signals toward sparser, harder ones, over simultaneous combination, and should expect this staging to matter as much as reward choice itself. 
Raising the advancement threshold between stages is a low-cost way to further stabilize training, though its benefit is uneven across benchmarks.
Because diffusion models can consume their full compute budget with near-zero learning progress without any corresponding increase in runtime, we recommend monitoring reward variance over training, rather than training time alone, as an early indicator of this failure.

\section{Conclusion}
\label{section: Conclusion}

This paper presents an empirical study of reward design for RL-based code generation across autoregressive and diffusion language models.
Reward effectiveness depends more on model architecture than on any single reward function: execution and static checking rewards are the most consistently beneficial, while format and similarity rewards often underperform the unrewarded baseline.
Diffusion models are far more sensitive to reward choice, with naive combination frequently causing performance collapse even when the same rewards help individually; progressive scheduling largely mitigates this. 
Training costs are also asymmetric: autoregressive models pay a wall-clock premium for execution-based supervision, while diffusion models can consume compute with near-zero learning progress under certain configurations. 
Overall, there is no single, universally optimal RL recipe for code generation: effective reward design depends on model architecture, combination and scheduling strategy, and the cost–performance trade-off.

\section{Limitations}

\paragraph{Training scope.} We train on the hard partition of AceCode-87k, since this is the setting where execution-based reward is sparsest, and reward design is therefore expected to matter most; training on easier problems would likely mask differences between reward functions, since execution reward alone is already dense when success rates are high.
This choice, however, means our main claims about how reward benefit interacts with task difficulty are inferred from evaluation-time differences across benchmarks of varying difficulty, rather than directly tested through difficulty-stratified training.
To partially address this, we report additional results from training on the easy and medium partitions in Appendix~\ref{appendix: Additional Experiment}, though these are not the focus of our main analysis.


\paragraph{RL algorithm choice.} We use a single RL algorithm, GRPO, throughout, so it is unclear whether the instabilities we attribute to reward design and architecture are specific to this algorithm or would generalize to other RL methods, such as PPO-based approaches.


\paragraph{Scale and evaluation scope.} Our four base models span a narrow 7--8B parameter range, so the architecture effects we identify, particularly diffusion models' heightened reward sensitivity and the zero-variance failure mode, may not hold at substantially smaller or larger scales.
Our evaluation is limited to Python and to relatively short, self-contained problems (HumanEval(+), MBPP(+)) plus LiveCodeBench; results may not transfer to other languages or to longer, more compositional software-engineering tasks.
We report pass@1 accuracy throughout, which does not capture code efficiency, readability, or maintainability, so reward functions that appear beneficial or harmful under pass@1 could rank differently under these other measures of code quality.

\section{Ethical Considerations}
\label{Ethical Considerations}

This work studies reward design for RL-based training of code generation models using publicly released base models and an existing benchmark dataset; it does not collect new human data, involve human subjects, or introduce new models or datasets for public release. 
We nonetheless note several risks relevant to the broader line of work this paper contributes to.

\paragraph{Downstream misuse of improved code generation.} Reward designs that make RL training more effective, particularly the execution and static-checking rewards we identify as most reliable, could accelerate the deployment of stronger code-generation systems.
Such systems carry the same dual-use risks as code LLMs generally: they could be used to generate malicious code (e.g., malware, exploits) as readily as benign code, and our results do not evaluate or filter for this.
We did not observe or specifically test for such behavior in this study, since all evaluation is conducted on standard, benign programming benchmarks.

\paragraph{Reliability and over-trust in generated code.} None of the reward functions we study, including execution-based reward, guarantee correctness beyond the specific test cases used for training or evaluation; a model optimized to pass benchmark tests may still produce code with latent bugs, security vulnerabilities, or behavior that generalizes poorly outside the benchmark distribution.
Practitioners adopting the reward designs we recommend should not treat improved benchmark pass rates as a substitute for standard software testing, review, and security auditing practices.

\paragraph{Environmental cost of RL training.} RL training for code generation, particularly with execution-based rewards, is computationally expensive: Section~\ref{section: Results} shows execution-based reward nearly doubling training time for autoregressive models, and our full experimental suite required 20+ GRPO training runs across four 7--8B models on a single H200 GPU (Appendix~\ref{appendix: Training}).
This computational cost carries a corresponding energy and carbon footprint. We view our training-efficiency analysis (Experiment 4) as a partial mitigation in this direction, since it is intended to help practitioners choose reward configurations that avoid wasted compute, particularly the zero-variance failure mode we identify for diffusion models, in which substantial compute can be spent with no corresponding learning progress.

\section*{Acknowledgments}
This work was supported by UK Research and Innovation [grant number EP/S023356/1], in the UKRI Centre for Doctoral Training in Safe and Trusted Artificial Intelligence (\url{https://www.safeandtrustedai.org}).
Part of this work was carried out during an internship at Elm Company.

\newpage
\bibliography{reference}

\newpage
\appendix

\section{Preliminary}
\label{appendix: Preliminary}

\subsection{AR Language Model}

An AR language model performs inference by generating tokens sequentially from left to right.
Given a prompt $\mathbf{p}$ that describes the coding task, the model produces a completion sequence $\mathbf{c} = (c_1, c_2, \ldots, c_L)$, where each token $c_i$ belongs to the vocabulary $\mathcal{V}$ or a special end-of-sequence token \texttt{EOS}.
The generation process is naturally formulated as a policy parameterized by weights $\theta$:
\begin{equation}
\pi_\theta(\cdot \mid \mathbf{p}, \mathbf{c}_{<i})
\;\triangleq\;
p_\theta(c_i \mid \mathbf{p}, \mathbf{c}_{<i}), 
\quad i \in [1, L],
\label{eq:ar_policy_def}
\end{equation}
where $\mathbf{c}_{<i} = (c_1, \ldots, c_{i-1})$ denotes the previously generated prefix. Under this factorization, the probability of generating the full completion $\mathbf{c}$ is
\begin{equation}
\pi_\theta(\mathbf{c} \mid \mathbf{p})
=
\prod_{i=1}^{L}
p_\theta(c_i \mid \mathbf{p}, \mathbf{c}_{<i}).
\label{eq:ar_sequence_prob}
\end{equation}

In code generation, the user provides a prompt $\mathbf{p}$ that specifies the programming problem, input-output requirements, and other task constraints.
The AR LM then generates a completion $\mathbf{c}$ token by token until it reaches \texttt{EOS} or a predefined maximum length.
The final code $\mathbf{y}$ can be extracted from the completion using a deterministic program such as regular-expression matching or code-block parsing, i.e.,
\begin{equation}
    \mathbf{y} = \text{Extract}(\mathbf{c}).
    \label{eq:ar_extract}
\end{equation}
The extracted code may remove irrelevant tokens such as natural-language explanations, reasoning traces, Markdown delimiters, or padding. 
We denote the human-annotated ground-truth code solution as $\mathbf{y}^*$. 

\subsection{Diffusion Language Model}

A masked DLM performs inference by gradually denoising a masked input sequence. Let $\mathbf{x}_T$ be the DLM's input sequence, where each element may be an unmasked token from the vocabulary $\mathcal{V}$ or a special \texttt{MASK} token. The final output of DLM decoding is $\mathbf{x}_0$ where all tokens belong to $\mathcal{V}$ and are therefore unmasked. The $\mathbf{x}_t, 0<t<T$ represent the intermediate sequences with partial masking. The DLM is a policy parameterized by weights $\theta$ that captures the unmasking process
\begin{equation}
\pi_\theta(\cdot \mid \mathbf{x}_t, t) \;\triangleq\; p_\theta(\mathbf{x}_{t-1}\mid \mathbf{x}_t, t), t \in [1, T]
\label{eq:policy_def}
\end{equation}
In code generation, the user provides a prompt $\mathbf{p}$ that describes the coding task.
The prompt is then concatenated with $L$ \texttt{MASK} tokens to form the DLM input, 
\begin{equation}
    \mathbf{x}_T \triangleq (\mathbf{p}, \texttt{MASK}^L)
    \label{eq:input_standard}
\end{equation} After DLM decoding, the final code $\mathbf{y}$ can be extracted from $\mathbf{x}_0$ by removing the prompt and other irrelevant tokens (such as the reasoning trace and the padding) using a deterministic program such as regular-expression matching, i.e., $\mathbf{y} = \text{Extract}(\mathbf{x}_0)$.
We denote the human-annotated ground-truth code solution as $\mathbf{y}^*$.

\section{Models}
\label{appendix: Models}

As shown in Table~\ref{tab:models}, we perform RL training on four instruction-tuned 7--8B code models
spanning two decoding paradigms: two AR models,
\texttt{Qwen/Qwen2.5-Coder-7B-Instruct} and
\texttt{ByteDance-Seed/Seed-Coder-8B-Instruct}, and two diffusion language
models, \texttt{apple/DiffuCoder-7B-Instruct} and
\texttt{Dream-org/Dream-Coder-v0-Instruct-7B}.
All four are loaded in
\texttt{bfloat16}.

\begin{table}[h!]
\centering
\small
\caption{Models used in our study.}
\label{tab:models}
\resizebox{\linewidth}{!}{
\begin{tabular}{l l r r}
\toprule
\textbf{Model} & \textbf{Paradigm} & \textbf{Size} & \textbf{Context Length} \\
\midrule
Qwen2.5-Coder & Autoregressive LM & 7B & 128K \\

Seed-Coder & Autoregressive LM & 8B & 32K \\

DiffuCoder & Diffusion LM & 7B & 128K \\

Dream-Coder & Diffusion LM & 7B & 32K \\

\bottomrule
\end{tabular}
}
\end{table}

\section{Reward}
\label{appendix: Reward}

As shown in Table \ref{table: comparison of the reward functions}, we design a multi-component reward that mirrors a practical code-generation pipeline:
(i) \emph{format} (whether code can be successfully extracted),
(ii) \emph{syntax} (whether the code is parsable),
(iii) \emph{similarity} (similarity with ground truth solutions),
(iv) \emph{static checking} (defect signals from static analysis), and
(v) \emph{execution} (test case pass rate).

\begin{table*}[h!]
\caption{Reward functions used in our paper. 
We summarize each reward by its requirements (whether it needs code execution, ground-truth solutions, or a static checker) and its properties (whether it provides a dense training signal, whether it is code-structure aware, and its relative time cost).}

\vspace{0mm}

\centering
\resizebox{\linewidth}{!}{
\begin{tabular}{l c c c | c c c}
\toprule
\multirow{2}{*}{Reward} & \multicolumn{3}{c|}{Requirements} & \multicolumn{3}{c}{Properties} \\
\cmidrule{2-7}
 & Code Execution & Ground Truth & Static Checker & Dense Design & Code Structure Aware & Time Cost \\
\midrule
Format & \xmark & \xmark & \xmark & \xmark & \xmark & Low \\ 
Syntax & \xmark & \xmark & \xmark & \xmark & \xmark & Low\\
Similarity & \xmark & \cmark & \xmark & \cmark & \cmark & Low\\
Static Checking & \xmark & \xmark & \cmark & \cmark & \xmark & Low\\
Execution & \cmark & \xmark & \xmark & \xmark & \cmark & High\\

\bottomrule
\end{tabular}
}
\label{table: comparison of the reward functions}
\end{table*}

\paragraph{Format reward}
The format reward targets the code extraction stage, and it seeks to ensure that the code $\mathbf{y}$ can be successfully extracted from the model output $\mathbf{x}_0$ without error. One simple implementation is to check if the extracted code is empty: 
\begin{equation}
r_{\mathrm{fmt}}(\mathbf{x}_0) \;=\; \mathbb{I}\!\left[\,\text{Extract}(\mathbf{x}_0)\neq \varnothing\,\right],
\label{eq:fmt}
\end{equation}
where $\mathbb{I}[\cdot]$ is the indicator function. The format reward does not improve the model's code generation capability by itself; rather it sets a basic requirement for the model.

\paragraph{Syntax reward}
Given the extracted code $\mathbf{y}$, the syntax reward checks whether it is syntactically well-formed by attempting to parse $\mathbf{y}$ into an Abstract Syntax Tree (AST), a tree-structured representation that captures the program's grammatical structure (e.g., expressions, statements, and declarations) while abstracting away surface details such as whitespace and formatting.
As an illustrative example, in Python this can be implemented by invoking \texttt{ast.parse}\footnote{\url{https://docs.python.org/3/library/ast.html}}.
Let $\mathrm{Parse}(\mathbf{y})$ denote the event that the chosen language parser successfully produces an AST without error.
We define
\begin{equation}
r_{\mathrm{syn}}(\mathbf{y}) \;=\; \mathbb{I}\!\left[\,\mathrm{Parse}(\mathbf{y})\,\right].
\label{eq:syn}
\end{equation}


\paragraph{Static checking reward}
While the syntax reward in Eq.~\eqref{eq:syn} only verifies that the generated program can be parsed into an AST, it provides a binary signal and cannot distinguish between code that is merely syntactically valid and code that is well-formed, maintainable, and less likely to contain shallow defects.
In contrast, our static checking reward leverages a static analyzer that does not execute the code to provide graded feedback on a broader set of properties that often correlate with downstream functional correctness and robustness.
Concretely, we use \texttt{Pylint}\footnote{\url{https://pylint.readthedocs.io/en/stable/index.html}} as the static checker, which reports an overall score by aggregating findings across multiple categories, including \texttt{Fatal}, \texttt{Error}, \texttt{Warning}, \texttt{Convention}, and \texttt{Refactor}.
These categories go beyond grammatical well-formedness and capture issues such as undefined names, unreachable code, inconsistent control flow, suspicious redefinitions, unused variables/imports, and style and complexity signals that reflect code quality.

Formally, given extracted code $\mathbf{y}$, let $s_{\mathrm{pylint}}(\mathbf{y})\in[0,10]$ denote the overall Pylint score computed under a \emph{fixed} Pylint version and a fixed ruleset (to ensure reward stationarity throughout RL training).
We convert it into a normalized reward in $[0,1]$ as
\begin{equation}
r_{\mathrm{pylint}}(\mathbf{y}) \;=\; \frac{s_{\mathrm{pylint}}(\mathbf{y})}{10}.
\label{eq:lint}
\end{equation}
The $r_{\mathrm{pylint}}$ yields a denser learning signal by capturing diverse aspects of the code quality.
In our training pipeline, we compute $r_{\mathrm{pylint}}$ only after successful code extraction and parsing, i.e., we gate static checking on $r_{\mathrm{syn}}(\mathbf{y})=1$, since Pylint analysis is unreliable when the program is not syntactically well-formed.


\paragraph{Similarity reward}
This reward measures how similar the generated program $\mathbf{y}$ is to the ground truth solution $\mathbf{y}^*$, from both syntactic and structural aspects.

\textit{Syntactic similarity.}
We tokenize each code and represent the token sequence as a TF--IDF vector over token $n$-grams with $n\in\{1,2,3\}$.
Let $\phi_{\text{tfidf}}(\cdot)$ denote this TF--IDF embedding \cite{abubakar2022sentiment}.
The syntactic similarity to reference $\mathbf{y}^*$ is computed by cosine similarity:
\begin{equation}
s^{\text{syn}}_i \;=\; \cos\!\Big(\phi_{\text{tfidf}}(\mathbf{y}),\, \phi_{\text{tfidf}}(\mathbf{y}^*)\Big)
\;\in\; [0,1].
\label{eq:syntax_sim}
\end{equation}

\textit{Structural similarity.}
We extract a sparse AST feature vector for each code.
Given a successfully parsed AST, we traverse the tree and extract features \cite{hu2022treecen}:
(i) \emph{node-type counts}, where each AST node type contributes a feature \texttt{node\_<Type>} equal to its frequency;
(ii) \emph{parent--child edge patterns}, where each observed parent--child type pair contributes a feature \texttt{edge\_<Parent>-><Child>} equal to its frequency;
and (iii) a \emph{depth} feature \texttt{max\_depth}, defined as the maximum depth reached during traversal.
This produces a sparse dictionary of real-valued counts for each program.
We then vectorize it to ensure aligned dimensions across the candidate $\mathbf{y}$ and reference $\mathbf{y}^*$: all keys observed in the set $\{\mathbf{y}\}\cup\{\mathbf{y}^*\}$ define a shared feature space, and missing keys are treated as zeros.
Let $\phi_{\text{ast}}(\cdot)$ denote the resulting feature embedding (vectorized with a dictionary vectorizer to ensure aligned dimensions across $\mathbf{y}$ and $\mathbf{y}^*$).
The structural similarity to reference $\mathbf{y}^*$ is again computed via cosine similarity:
\begin{equation}
s^{\text{ast}}_i \;=\; \cos\!\Big(\phi_{\text{ast}}(\mathbf{y}),\, \phi_{\text{ast}}(\mathbf{y}^*)\Big)
\;\in\; [0,1].
\label{eq:ast_sim}
\end{equation}

\textit{Combined similarity.}
For each reference solution, we combine syntactic and structural similarities with a convex weight $\alpha \in [0,1]$ (here we use 0.3 based on empirical practice):
\begin{equation}
s_i \;=\; \alpha\, s^{\text{syn}}_i \;+\; (1-\alpha)\, s^{\text{ast}}_i .
\label{eq:combined_sim}
\end{equation}

\paragraph{Execution reward}
The execution reward, also known as pass rate, directly measures functional correctness via \textbf{test execution}.
It is one of the most widely used rewards in RL for code \cite{gong2025diffucoder, xie2025dream}.
Let $\mathcal{C}$ be the set of test cases for prompt $\mathbf{p}$, and let $\mathrm{Pass}(\mathbf{y},\tau)$ indicate whether code $\mathbf{y}$ passes test case $\tau\in\mathcal{C}$ under sandboxing and timeouts.
We define test pass rate as
\begin{equation}
r_{\mathrm{exec}}(\mathbf{y},\mathcal{C}) \;=\; \frac{1}{|\mathcal{C}|}\sum_{\tau\in\mathcal{C}} \mathbb{I}\!\left[\,\mathrm{Pass}(\mathbf{y},\tau)\,\right].
\label{eq:exec}
\end{equation}

\section{Training}
\label{appendix: Training}

This appendix reports the exact training configuration used for Section~\ref{section: Results}.

\paragraph{Algorithm and dataset.} All runs use Group Relative Policy Optimization (GRPO), training on the hard split of AceCode-87k with reference solutions and unit tests (1{,}438 problems), capped at 719 optimizer steps.
Each run applies the five candidate reward functions under one of the three regimes studied in this paper: individually, in combination, or under progressive scheduling. 
The five rewards span three cost tiers: \emph{lightweight} (format, syntax correctness, computed by pattern or AST checks on the raw completion), \emph{static analysis} (static checking, similarity to the reference solution, neither of which executes the candidate code), and \emph{execution-based} (test pass rate, which executes the candidate against its unit tests).
Execution-based reward runs candidate code in a sandboxed executor; generated code is truncated to a fixed maximum length before evaluation, and syntax/format rewards similarly cache only a bounded prefix of the completion.

\paragraph{Optimization hyperparameters.} The optimizer, learning-rate schedule, GRPO-specific coefficients, and precision/logging settings are held fixed across all runs, including the individual-reward, combined-reward, and progressive-scheduling regimes.
All models are trained with AdamW ($\beta_1{=}0.9$, $\beta_2{=}0.99$, $\epsilon{=}10^{-8}$) at a learning rate of $1\times 10^{-6}$ under a cosine schedule with a minimum-LR floor (10\% of peak) and a short warmup ratio ($10^{-4}$, no explicit warmup steps). Weight decay is 0.1 and gradients are clipped to a max norm of 0.2.
Each GRPO update performs 2 inner iterations over the sampled batch, with a KL penalty coefficient $\beta{=}0.01$ against a reference model that is resynchronized to the policy every 64 steps with mixup coefficient $\alpha{=}0.6$.
All runs use bfloat16 precision with gradient checkpointing enabled, a fixed seed of 42, and train for up to 719 optimizer steps.

\paragraph{Architecture-specific settings.} Sequence-length budgets, sampling parameters, group size, and effective batch size are \emph{not} identical across architectures.
All models, AR and diffusion alike, are trained on a single H200 GPU.

The two diffusion models, DiffuCoder-7B and Dream-Coder-7B, are trained with a per-device train batch size of 5 and gradient accumulation of 2 steps, sampling 10 generations per prompt at temperature 1.0 (top-$p{=}1.0$, top-$k{=}50$) with a max prompt length of 200 tokens and a max completion length of 256 tokens.
Diffusion decoding runs a fixed 256-step denoising schedule with random prompt/completion masking.
They use a wider PPO-style clip range ($\epsilon{=}0.5$), raw (unscaled) rewards, and a plain assistant-style system prompt.
The two AR models, Qwen2.5-Coder-7B and Seed-Coder-8B, are trained with a proportionally larger per-device train batch size of 16 (gradient accumulation again 2 steps), sampling 8 generations per prompt at temperature 1.0 (top-$p{=}1.0$, top-$k{=}50$) with a max prompt length of 1024 tokens and an $8\times$ longer max completion length of 2048 tokens, reflecting their variable-length autoregressive decoding.
They use the more conventional clip range $\epsilon{=}0.2$ with reward scaling enabled, and a reasoning-style system prompt that instructs the model to think before answering.

\section{Evaluation}
\label{appendix: Evaluation}

We evaluate trained checkpoints on three held-out code-generation benchmarks: HumanEval(+) and MBPP(+) via EvalPlus, and LiveCodeBench (release v6, code-generation scenario).
All three are scored under a single-sample, greedy-decoding protocol (temperature 0, one completion per problem) and report pass@1 as the primary metric, with no additional fine-tuning beyond the GRPO checkpoint being evaluated.
As with training, evaluation uses separate procedures for the AR models (Qwen2.5-Coder-7B, Seed-Coder-8B) and the diffusion models (DiffuCoder-7B, Dream-Coder-7B), since the two families require different generation backends; both are run on a single GPU.

\paragraph{AR models.} For HumanEval/MBPP, evaluation calls EvalPlus's vLLM backend directly on each checkpoint, which handles generation, sanitization, and scoring internally and reports separate pass@1 for the base test suite and the augmented EvalPlus test suite.
For LiveCodeBench, evaluation uses the standard LiveCodeBench runner in the code-generation scenario against release v6, with the corresponding base model identifier used to resolve the correct tokenizer and chat template; generation uses greedy decoding with one completion per problem, and scoring is performed in the same run.

\paragraph{Diffusion models.} Because neither EvalPlus's vLLM backend nor the standard LiveCodeBench runner natively supports diffusion-LM decoding, the diffusion evaluation pipeline instead routes through custom generation utilities before handing off to the same scoring tools used for the AR models.
For HumanEval/MBPP, raw completions are generated per problem for a given checkpoint and dataset, then post-processed with EvalPlus's sanitization step to extract the code block, and finally scored with the same EvalPlus scoring command used on the AR side, so the resulting pass@1 numbers are directly comparable across architectures.
For LiveCodeBench, an analogous generation utility produces one greedy completion per problem against release v6; candidate programs are executed and scored in parallel with a fixed per-problem execution timeout, and pass@1 is aggregated into the same summary format used for the AR models.


\begin{table*}[h!]
\centering
\small
\setlength{\tabcolsep}{6pt}
\setlength{\extrarowheight}{2pt}
\setlength{\aboverulesep}{0pt}
\setlength{\belowrulesep}{0pt}
\caption{Effect of individual reward components on code generation benchmarks across difficulty partitions (Easy, Medium, Hard). The base model (grey row) is shown once per model as ``--'' for reference. \colorbox{highlight}{Light green} cells indicate a reward outperforming the base model on that metric; \textbf{bold} indicates the best value in each column, within each difficulty block.}
\label{tab:model_evaluation}
\begin{tabular}{@{}lllrrrrr}
\toprule
\textbf{Model} & \textbf{Difficulty} & \textbf{Reward} & \textbf{HumanEval} & \textbf{HumanEval+} & \textbf{MBPP} & \textbf{MBPP+} & \textbf{LiveCodeBench} \\
\midrule
\multirow{16}{*}{Qwen2.5-Coder}
 & \cellcolor{gray!20}-- & \cellcolor{gray!20}-- & \cellcolor{gray!20}0.884 & \cellcolor{gray!20}0.841 & \cellcolor{gray!20}0.825 & \cellcolor{gray!20}0.709 & \cellcolor{gray!20}0.297 \\
\cmidrule{2-8}
 & \multirow{5}{*}{Easy}
   & Format          & \cellcolor{highlight}\textbf{0.915} & 0.841 & \cellcolor{highlight}0.831 & 0.709 & \cellcolor{highlight}\textbf{0.302} \\
 & & Syntax          & \cellcolor{highlight}\textbf{0.915} & \cellcolor{highlight}0.848 & \cellcolor{highlight}0.849 & \cellcolor{highlight}\textbf{0.722} & \cellcolor{highlight}0.300 \\
 & & Similarity      & \cellcolor{highlight}0.909 & \cellcolor{highlight}\textbf{0.854} & \cellcolor{highlight}0.828 & 0.701 & 0.297 \\
 & & Static Checking & 0.884 & 0.829 & \cellcolor{highlight}\textbf{0.884} & \cellcolor{highlight}0.720 & 0.295 \\
 & & Execution       & \cellcolor{highlight}0.902 & 0.841 & \cellcolor{highlight}0.841 & 0.709 & \cellcolor{highlight}0.301 \\
\cmidrule{2-8}
 & \multirow{5}{*}{Medium}
   & Format          & \cellcolor{highlight}\textbf{0.896} & 0.841 & \cellcolor{highlight}0.841 & \cellcolor{highlight}0.722 & 0.296 \\
 & & Syntax          & 0.878 & 0.817 & 0.820 & 0.701 & 0.285 \\
 & & Similarity      & \cellcolor{highlight}0.890 & \cellcolor{highlight}\textbf{0.848} & \cellcolor{highlight}0.841 & \cellcolor{highlight}0.714 & 0.295 \\
 & & Static Checking & 0.872 & 0.811 & \cellcolor{highlight}0.833 & 0.685 & 0.292 \\
 & & Execution       & \cellcolor{highlight}0.890 & 0.835 & \cellcolor{highlight}\textbf{0.862} & \cellcolor{highlight}\textbf{0.728} & \cellcolor{highlight}\textbf{0.298} \\
\cmidrule{2-8}
 & \multirow{5}{*}{Hard}
   & Format          & 0.884 & 0.811 & 0.825 & 0.704 & \cellcolor{highlight}0.298 \\
 & & Syntax          & \cellcolor{highlight}0.896 & 0.817 & \cellcolor{highlight}0.844 & \cellcolor{highlight}0.717 & 0.295 \\
 & & Similarity      & 0.866 & 0.799 & \cellcolor{highlight}0.839 & 0.709 & 0.280 \\
 & & Static Checking & 0.854 & 0.787 & \cellcolor{highlight}\textbf{0.862} & 0.698 & \cellcolor{highlight}\textbf{0.300} \\
 & & Execution       & \cellcolor{highlight}\textbf{0.915} & \cellcolor{highlight}\textbf{0.854} & \cellcolor{highlight}0.852 & \cellcolor{highlight}\textbf{0.720} & 0.291 \\
\midrule
\multirow{16}{*}{DiffuCoder}
 & \cellcolor{gray!20}-- & \cellcolor{gray!20}-- & \cellcolor{gray!20}0.652 & \cellcolor{gray!20}0.616 & \cellcolor{gray!20}0.701 & \cellcolor{gray!20}0.590 & \cellcolor{gray!20}0.134 \\
\cmidrule{2-8}
 & \multirow{5}{*}{Easy}
   & Format          & \cellcolor{highlight}\textbf{0.677} & \cellcolor{highlight}\textbf{0.622} & 0.696 & 0.590 & \cellcolor{highlight}\textbf{0.135} \\
 & & Syntax          & 0.591 & 0.537 & 0.685 & 0.579 & 0.128 \\
 & & Similarity      & 0.598 & 0.555 & 0.690 & \cellcolor{highlight}0.601 & 0.126 \\
 & & Static Checking & \cellcolor{highlight}0.659 & 0.604 & 0.690 & 0.585 & 0.115 \\
 & & Execution       & 0.573 & 0.524 & \cellcolor{highlight}\textbf{0.709} & \cellcolor{highlight}\textbf{0.611} & 0.123 \\
\cmidrule{2-8}
 & \multirow{5}{*}{Medium}
   & Format          & 0.585 & 0.549 & 0.696 & \cellcolor{highlight}0.593 & 0.112 \\
 & & Syntax          & 0.573 & 0.530 & 0.680 & 0.571 & 0.089 \\
 & & Similarity      & 0.622 & \textbf{0.573} & 0.690 & 0.585 & 0.118 \\
 & & Static Checking & \textbf{0.628} & 0.561 & \cellcolor{highlight}0.714 & \cellcolor{highlight}0.601 & \textbf{0.122} \\
 & & Execution       & 0.598 & 0.561 & \cellcolor{highlight}\textbf{0.720} & \cellcolor{highlight}\textbf{0.606} & \textbf{0.122} \\
\cmidrule{2-8}
 & \multirow{5}{*}{Hard}
   & Format          & 0.622 & 0.567 & \cellcolor{highlight}0.706 & \cellcolor{highlight}0.595 & 0.126 \\
 & & Syntax          & 0.494 & 0.445 & 0.651 & 0.561 & 0.100 \\
 & & Similarity      & \cellcolor{highlight}\textbf{0.689} & \cellcolor{highlight}\textbf{0.634} & 0.688 & 0.577 & 0.071 \\
 & & Static Checking & \cellcolor{highlight}0.677 & 0.610 & \cellcolor{highlight}\textbf{0.725} & \cellcolor{highlight}0.608 & \cellcolor{highlight}\textbf{0.155} \\
 & & Execution       & 0.591 & 0.549 & \cellcolor{highlight}0.720 & \cellcolor{highlight}\textbf{0.627} & 0.123 \\
\bottomrule
\end{tabular}
\end{table*}

\section{Additional Experiment}
\label{appendix: Additional Experiment}

\subsection{How sensitive is each reward to training data difficulty?}

For the AR model, four of the five rewards show a broadly declining benefit as training difficulty rises from easy to hard. Format reward's HumanEval gain shrinks from +0.031 over baseline at easy (0.915) to 0 at hard (0.884, matching the base model exactly).
Similarity reward follows a similar pattern on HumanEval, declining steadily from 0.909 at easy to 0.890 at medium and 0.866 at hard, and also reaches its lowest LiveCodeBench value at hard (0.280).
Static checking shows the same downward trend, moving from a tie with baseline at easy (0.884 HumanEval) to a $-0.030$ deficit at hard (0.854), with MBPP+ falling from 0.720 to 0.698 over the same range.
This suggests that, for Qwen2.5-Coder, several reward signals lose discriminative power as tasks become harder, plausibly because similarity- and static-checking-based signals become less correlated with true functional correctness once superficially clean or textually similar solutions are more likely to still be incorrect.

Execution reward is the clearest exception to this pattern on HumanEval, staying above baseline (0.884) at all three tiers, though non-monotonically: 0.902 at easy, a slight dip to 0.890 at medium, then a rise to 0.915 at hard.
This robustness does not extend to every benchmark: on LiveCodeBench, execution's margin over baseline (0.297) narrows as difficulty increases, from 0.301 at easy to 0.298 at medium to 0.291 at hard, falling below baseline for the first time.
This indicates that difficulty robustness is itself benchmark-dependent: a reward that stays reliable on benchmarks similar in distribution to the training data (HumanEval, MBPP) can still lose robustness on a more out-of-distribution benchmark like LiveCodeBench.
Syntax reward shows a related but noisier pattern, scoring well at easy (0.915) and recovering at hard (0.896), but dipping below the base model at medium (0.878 vs.\ 0.884), making it the least predictable reward for this model across difficulty levels.

The diffusion model does not follow a single difficulty trend across rewards; different rewards respond to increasing difficulty in opposite directions.
Syntax reward degrades sharply and nearly monotonically, from an already below-baseline 0.591 HumanEval at easy to 0.494 at hard, the lowest HumanEval score recorded anywhere in this study, with LiveCodeBench following a similar downward path (0.128 $\to$ 0.089 $\to$ 0.100).
Format reward is similarly fragile to difficulty, falling from above baseline at easy (0.677 vs.\ 0.652) to below baseline at both medium (0.585) and hard (0.622).

In contrast, static checking improves for DiffuCoder as difficulty increases on most measures: its HumanEval score rises from 0.659 at easy to 0.677 at hard (dipping slightly below baseline at medium, 0.628 vs.\ 0.652), and its LiveCodeBench score climbs substantially, from 0.115 at easy to 0.155 at hard, the best LiveCodeBench result for DiffuCoder in Table~\ref{tab:model_evaluation}.
Similarity reward shows the most striking divergence: on HumanEval it becomes more effective as difficulty increases, reaching 0.689 at hard, above the 0.652 baseline and the best hard-difficulty result for this model, while MBPP stays roughly flat (0.690 $\to$ 0.690 $\to$ 0.688) and MBPP+ declines (0.601 $\to$ 0.585 $\to$ 0.577).
On LiveCodeBench, however, similarity moves in the opposite direction, falling from 0.126 at easy to 0.071 at hard, the lowest LiveCodeBench score in the entire table. 
This suggests that training DiffuCoder with a similarity-based reward on harder problems may sharpen its ability to solve problems structurally close to the training distribution while narrowing its generalization to LiveCodeBench's more novel problems, a difficulty-dependent trade-off that would not be visible from a single held-out benchmark alone.

\section{Threats to Validity}
\label{section: Threats to Validity}

\paragraph{Internal Validity.}
Our progressive-scheduling thresholds (0.5, 0.8) and reward ordering (format $\to$ syntax $\to$ static checking $\to$ execution) are chosen a priori rather than tuned per model, so the reported gains reflect one fixed, hand-designed schedule rather than the ceiling achievable with model-specific tuning.
Similarly, all runs use matched hyperparameters across models and reward configurations to ensure a fair comparison, but this may under-serve some model-reward combinations relative to what individually tuned settings would achieve, which could conflate genuine architecture sensitivity with hyperparameter mismatch.

\paragraph{External Validity.}
Our four base models span a narrow 7--8B parameter range, so the architecture effects we identify, particularly diffusion models' heightened reward sensitivity and susceptibility to the zero-variance failure mode, may not generalize to substantially smaller or larger models.
We evaluate on HumanEval(+), MBPP(+), and LiveCodeBench, all Python-centric, relatively short-form programming benchmarks, so our findings may not transfer to other languages, longer or more compositional software-engineering tasks, or repository-level code generation.
We report pass@1 accuracy as our primary metric, which does not capture code efficiency, readability, or maintainability, so reward functions that appear beneficial or harmful under pass@1 might rank differently under these alternative measures of code quality.
Finally, despite LiveCodeBench's design to reduce train--test contamination, we cannot fully rule out residual overlap between AceCode-87k and our evaluation benchmarks, which could inflate absolute performance estimates across all reward configurations alike.




\end{document}